\documentclass[twocolumn,aps,prb]{revtex4}
\usepackage[dvips]{graphicx}
\usepackage{color,array,dcolumn}
\usepackage{amsmath}
\usepackage{amssymb}

\begin{document}

\title{Van der Waals Interactions in DFT using Wannier Functions: improved 
$C_6$ and $C_3$ coefficients by a new approach}
\author{A. Ambrosetti}
\email{ambroset@pd.infn.it}
\affiliation{Dipartimento di Fisica, University of Padova, via Marzolo 8, I--35131, 
Padova, Italy,
and DEMOCRITOS National Simulation Center, Trieste, Italy}
\author{P.L. Silvestrelli}
\affiliation{Dipartimento di Fisica, University of Padova, via Marzolo 8, I--35131, 
Padova, Italy,
and DEMOCRITOS National Simulation Center, Trieste, Italy}

\begin{abstract}
\date{\today}
A new implementation is proposed for including van der Waals interactions
in Density Functional Theory using the Maximally-Localized Wannier 
functions. With respect to the previous DFT/vdW-WF method, the
present DFT/vdW-WF2 approach, which is based on the simpler
London expression and takes into account the intrafragment overlap
of the localized Wannier functions, leads to a considerable improvement
in the evaluation of the $C_6$ van der Waals coefficients, as
shown by the application to a set of selected dimers.
Preliminary results on Ar on graphite and Ne on the Cu(111) metal
surface suggest that also the $C_3$ coefficients, characterizing
molecule-surfaces van der Waals interactions are better estimated
with the new scheme.  
\end{abstract}

\maketitle

An accurate description of ubiquitous, long-range van der Waals (vdW) 
interactions is crucial for characterizing countless phenomena,
belonging to such diverse fields as solid state and surface physics, 
chemistry and biology. For instance, vdW effects are responsible for the
stabilization of non covalently-bonded crystals and layered structures, 
play a major role in physisorption processes, and are know to affect 
several biological phenomena.
vdW interactions are due to long range-correlations, in particular, 
the leading $R^{-6}$ term is a consequence of correlated, instantaneous 
dipole fluctuations.
Density Functional Theory (DFT), thanks to its favorable scaling properties, 
represents a popular, efficient and invaluable approach, that is also
applicable to extended systems where other ab initio schemes turn out
to be too computationally expensive. However, standard DFT schemes
only provide a local or semilocal treatment of the electronic correlation,
so that they are unable to properly reproduce genuine vdW effects.\cite{Kohn} 

The simplest way to include vdW interactions in DFT is represented
by semiempirical methods,\cite{Grimme,Ortmann} where, typically,
an approximately derived $C_6/R^{-6}$ term is multiplied by a short-range 
damping function, with parameters tailored to the specific system considered.
Although such an approach is very efficient and often gives a substantial
improvement with respect to a standard DFT method, nonetheless its accuracy
is difficult to asses in advance and lacks of transferability 
(for instance, changes in atomic polarizabilities by changing the
atom environment are neglected). 
Clearly, better reliability, accuracy, and transferability can be in principle
achieved by adopting schemes where vdW corrections are computed by
exploiting the knowledge of the electronic density distribution given
by DFT. In recent years several approaches have been indeed proposed
(for a recent review, see, for instance, ref. \onlinecite{Riley})
In order to circumvent the direct use of truly non-local DFT functionals, 
which are not easy to evaluate efficiently, some of these methods introduce
suitable partitioning schemes into separated interacting fragments,
either relying on effective atom-atom\cite{TS,johnson,dftd3}
or electronic orbital-orbital\cite{silvprl,silvmetodo,Sato,Mostofi}
pairwise $C_6/R^6$ terms.
Although these techniques are expected to be more reliable and transferable
than semiempirical approaches, in practice, most of them
use one or more parameters to be fitted using some reference database.

Here we describe and apply a new implementation of the DFT/vdW-WF 
method\cite{silvprl,silvsurf,silvmetodo,Mostofi} where 
electronic charge partitioning
is achieved using the Maximally-Localized Wannier Functions (MLWFs).
The MLWFs are obtained from a unitary transformation in the space of the
occupied Bloch states, by minimizing the total spread functional:\cite{wannier}
 
\begin{equation}
\Omega=\sum_n S^2_n=\sum_n \left( <w_n|r^2|w_n>-<w_n|\mathbf{r}|w_n>^2\right).
\end{equation}

The localization properties of the MLWFs are of particular interest for 
the implementation of an efficient
vdW correction scheme: in fact, the MLWFs represent a suitable basis set
to evaluate orbital-orbital vdW interaction terms. 
While in the original DFT/vdW-WF method the vdW energy correction for
two separate fragments was computed using the exchange-correlation
functional proposed by Andersson {\it et al.},\cite{andersson} our
novel version (DFT/vdW-WF2 method) is instead based on the simpler,
well known London's expression:\cite{london} basically, 
two interacting atoms, $A$ and $B$, are approximated by 
coupled harmonic oscillators
and the vdW energy is taken to be the change of the zero-point energy
of the coupled oscillations as the atoms approach; if only a single
excitation frequency is associated to each atom, $\omega_A$, $\omega_B$,
then

\begin{equation}
E^{London}_{vdW}=-\frac{3e^4}{2m^2}\frac{Z_A Z_B}{\omega_A \omega_B(\omega_A+\omega_B)}\frac{1}{R_{AB}^6}
\label{lond}
\end{equation}

where $Z_{A,B}$ is the total charge of A and B, and $R_{AB}$ is 
the distance between the two atoms ($e$ and $m$ are the electronic charge
and mass).
Now, adopting a simple classical theory of the atomic polarizability, 
the polarizability 
of an electronic shell of charge $eZ_i$ and mass $mZ_i$, tied to a heavy 
undeformable ion can be written as

\begin{equation}
\alpha_i\simeq \frac{Z_i e^2}{m\omega_i^2}\,.
\label{alfa}
\end{equation}

Then, given the direct relation between polarizability and atomic
volume,\cite{polvol} we assume that $\alpha_i\sim \gamma S_i^3$,
where $\gamma$ is a proportionality constant, so that the atomic volume is
expressed in terms of the MLWF spread, $S_i$.
Rewriting eq. \eqref{lond} in terms of the quantities defined above, 
one obtains an explicit expression (much simpler than the multidimensional
integrals involved in the Andersson functional\cite{andersson}) for
the $C_6$ vdW coefficient: 

\begin{equation}
C_{6}^{AB}=\frac{3}{2}\frac{\sqrt{Z_A Z_B}S_A^3 S_B^3 \gamma^{3/2}}
{(\sqrt{Z_B}S_A^{3/2}+\sqrt{Z_A}S_B^{3/2})}\,.
\label{c6}
\end{equation}

The constant $\gamma$ can then be set up by imposing that the exact value for 
the H atom polarizability
($\alpha_H=$0.866 a.u.) is obtained (of course, in the H case, one
knows the exact analytical spread, $S_i=S_H=\sqrt{3}$ a.u.).
Note that, by expressing the ''atomic'' volume as a function of $S_i$ we
actually implicitly switch from an atom-atom to an orbital-orbital approach. 

In order to achieve a better accuracy, one must properly deal with
{\it intrafragment} MLWF overlap: in fact, the DFT/vdW-WF method is strictly
valid for nonoverlapping fragments only; now, while the overlap between the
MLWFs relative to separated fragments is usually negligible for all the
fragment separation distances of interest, the same is not true for the 
MLWFs belonging to the same fragment, which are often characterized by
a significant overlap. This overlap affects the effective orbital volume,
the polarizability, and the excitation 
frequency (see eq. \eqref{alfa}), thus leading to a quantitative effect on the
value of the $C_6$ coefficient. 
We take into account the effective change in volume due to intrafragment 
MLWF overlap by introducing a suitable reduction factor $\xi$
obtained by interpolating between the limiting cases of fully
overlapping and non-overlapping MLWFs.
In particular,
since in the present DFT/vdW-WF2 method the $i$-th MLWF is approximated 
with a homogeneous charged sphere of radius $S_i$, then the 
overlap among neighboring MLWFs can be evaluated as the geometrical
overlap among neighboring spheres. 
To derive the correct volume reweighting factor for dealing with overlap
effects, we first consider the limiting case of two pairs (one for
each fragment) of completely overlapping MLWFs, which would be, for instance,
applicable to two interacting He atoms if each MLWF just describes the
density distribution of a single electron; then we can evaluate
a single $C_6$ coefficient using eq. \eqref{c6} with $Z_{A,B}=2$,
so that:
 
\begin{equation}
C_{6}^{AB}=\frac{3}{2}\frac{\sqrt{2}S_A^3 S_B^3 \gamma^{3/2}}
{(S_A^{3/2}+S_B^{3/2})}.
\label{c62}
\end{equation}

Alternatively, the same expression can be obtained by considering the
sum of 4 identical pairwise contributions (with $Z=1$), by 
introducing a modification of the effective volume in such a way to
take the overlap into account and make the global interfragment
$C_6$ coefficient equivalent to that in eq. \eqref{c62}.
This is clearly accomplished by replacing $S_{i}^3$ in eq. \eqref{c6} with 
$\xi S_{i}^3$, where $\xi = 1/2$.
This procedure can be easily generalized to multiple 
overlaps, by weighting the overlapping
volume with the factor $n^{-1}$, where $n$ is the number of overlapping MLWFs.
Finally, by extending the approach to partial overlaps,  
we define the {\it free} volume of a set of MLWFs belonging to a given fragment
(in practice three-dimensional integrals are evaluated by numerical sums
introducing a suitable mesh in real space) as:

\begin{equation}
V_{free}=\int d\mathbf{r}\, w_{free}(\mathbf{r})
\simeq \Delta r \sum_l w_{free}(\mathbf{r}_l)
\end{equation}
where $w_{free}(\mathbf{r}_l)$ is equal to 1 if 
$|\mathbf{r}_l-\mathbf{r}_i|<S_i$ for at least one of the
fragment MLWFs, and is 0 otherwise.

The corresponding {\it effective} volume is instead given by
\begin{equation}
V_{eff}=\int d\mathbf{r}\, w_{eff}(\mathbf{r})
\simeq \Delta r \sum_l w_{eff}(\mathbf{r}_l)\,,
\end{equation}
where the new weighting function is defined as
$w_{eff}(\mathbf{r}_l)=w_{free}(\mathbf{r}_l)\cdot n_w(\mathbf{r}_l)^{-1}$,
with $n_w(\mathbf{r}_l)$ that is equal to the number of MLWFs
contemporarily satisfying the relation $|\mathbf{r}_l-\mathbf{r}_i|<S_i$.
Therefore, the non overlapping portions of the spheres (in
practice the corresponding mesh points) will be
associated to a weight factor 1, those belonging to two spheres to
a $1/2$ factor, and, in general, those belonging to $n$ spheres to
a $1/n$ factor. 
The average ratio between the effective volume and the free volume 
($V_{eff}/V_{free}$)
is then assigned to the factor $\xi$, appearing in eq. \eqref{c6eff}.
Although in principle the correction factor $\xi$
must be evaluated for each MLWF and the calculations must be repeated 
at different fragment-fragment separations, our tests show that,
in practice, if the fragments are rather homogeneous all the $\xi$ factors
are very similar, and if the spreads of the MLWFs do not change 
significantly in the range of the interfragment distances of
interest, the $\xi$'s remain essentially constant; clearly, 
exploiting this behavior leads to a significant reduction 
in the computational cost of accounting for the intrafragment overlap.
We therefore arrive at the following expression for the $C_6$ coefficient:

\begin{equation}
C_{6}^{AB}=\frac{3}{2}\frac{\sqrt{Z_A Z_B}\xi_A S_A^3 \xi_B S_B^3 \gamma^{3/2}}
{(\sqrt{Z_B\xi_A} S_A^{3/2}+\sqrt{Z_A\xi_B} S_B^{3/2})}\,,
\label{c6eff}
\end{equation}

where $\xi_{A,B}$ represents the ratio between the effective and the 
free volume associated to the $A$-th and $B$-th MLWF.
The need for a proper treatment of overlap effects has been also 
recently pointed out by Andrinopoulos {\it et al.},\cite{Mostofi} who however 
applied a correction only to very closely centred WFCs.

Finally, the vdW interaction energy is computed as:

\begin{equation}
E_{vdW}=-\sum_{i<j}f(R_{ij})\frac{C_6^{ij}}{R^6_{ij}}
\end{equation}
where $f(R_{ij})$ is a short-range damping function, which is
introduced not only to avoid the unphysical divergence of the
vdW correction at small fragment separations, but also
to eliminate double countings of correlation effects 
(in fact standard DFT approaches are able to describe short-range 
correlations); it is defined as:   
\begin{equation}
f(R_{ij})=\frac{1}{1+e^{-a(R_{ij}/R_s-1)}}\,.
\end{equation}

The parameter $R_s$ represents
the sum of the vdW radii $R_s=R_i^{vdW}+R_j^{vdW}$, with
(by adopting the same criterion chosen above for 
the $\gamma$ parameter)
\begin{equation}
R_i^{vdW}=R_H^{vdW}\frac{S_i}{\sqrt{3}}
\end{equation}
where $R_H^{vdW}$ is the literature\cite{Bondi} (1.20 \AA) vdW radius of 
the H atom, and, following Grimme {\it et al.},\cite{Grimme} 
$a \simeq 20$ (the results are almost
independent on the particular value of this parameter).
Although this damping function introduces a certain degree of empiricism
in the method, we stress that $a$
is the only ad-hoc parameter present in our approach, while all the others
are only determined by the basic information given by the MLWFs, namely 
from first principles calculations.

Calculations were performed using the CPMD code\cite{CPMD} and taking,
as the reference DFT GGA functional, 
both the PBE\cite{PBE} and revPBE\cite{revPBE} flavor:
PBE is chosen because it represents one of the most
popular GGA functionals for standard DFT simulations of condensed-matter
systems, while revPBE, which usually gives results close to those obtained by 
a pure Hartree-Fock approach, has been used both in our previous 
DFT/vdW-WF calculations and also in other vdW-corrected 
DFT studies.\cite{Langreth04,Grimme} 
As already pointed out elsewhere,\cite{silvsurf,ambr,Langreth04} 
vdW-corrected PBE 
calculations show a general tendency to overbinding (attributed to
an overestimate of the long-range part of the
exchange contribution), with equilibrium distances
in reasonable agreement with reference values, while instead 
vdW-corrected revPBE typically overestimates the equilibrium distances 
but gives better estimates for the binding energies.

We stress that the computational cost of the DFT/vdW-WF2 method,
although slightly increased with respect to that of the previous 
DFT/vdW-WF scheme, still represents a negligible additional cost if 
compared to that of a standard DFT calculation, thus satisfying the 
basic efficiency requirement.

In Table \ref{tabellac6} we report the $C_6$ coefficients computed
for a set of 18 dimolecular systems, where vdW interactions represent
the dominant (or at least a significant) contribution,
using our DFT/vdW-WF2 method to be compared to reference data. 
As can be seen, in most of the systems the $C_6$ coefficient value is
reduced and the overall performance
is much improved with respect to the previous DFT/vdW-WF approach:\cite{silvprl}
in fact the {\it mean relative error} (MRE) is decreased from 14.1 to 0.3 \%,
and from 16.6 to -0.3 \% with revPBE and PBE, respectively, while the
corresponding reductions in the {\it mean absolute relative error} (MARE) are
from 35.4 to 14.6 \% with revPBE and from 32.8 to 10.8 \% with PBE.
The effect is particularly apparent in rare-gas dimers and dimolecular 
complexes containing benzene, which are systems where the 
correction factor $\xi$ is important due to a significant overlap 
among MLWFs belonging to the same fragment. Clearly, one expects that
this correction will be also important in large molecules and also
in extended systems, characterized by relatively delocalized electronic
charge distributions, corresponding to large MLWF spreads.

Note that, the typical decrease of the $C_6$ coefficient values obtained
by the DFT/vdW-WF2 method does not necessarily lead to a reduction of
the vdW energy contribution; this is clearly due to the effect of
the adopted damping function, which determines the interplay between
the $C_6/R^{-6}$ vdW correction and standard DFT energy contributions.
As a result, the new DFT/vdW-WF2 scheme in general predicts 
slightly shorter equilibrium distances, in better agreement with reference
data than DFT/vdW-WF, while instead the binding energies exhibit
a behavior similar to that obtained by the DFT/vdW-WF approach,
basically determined by the underlying DFT GGA functional
(see above comment).
We also point out that, a much more accurate estimate of the
$C_6$ coefficients allows for a better description of the vdW interactions
even for interfragment distances far from the equilibrium values, which
is of particular relevance, both for the applications to large systems
and for Molecular Dynamics simulations.  

In order to test the applicability of the present DFT/vdW-WF2 method also 
to extended systems, which of course represent the most interesting 
application field because high-quality chemistry methods are too
computationally demanding, we considered both 
the adsorption of a single Ar atom on graphite and of a Ne atom
on the Cu(111) metal surface, which represent two typical physisorption 
processes.
In the case of Ar on graphite, calculations have been performed using
the same approach followed in ref. \onlinecite{ambr}, while for Ne on Cu(111) 
we have used the Quantum-ESPRESSO\cite{ESPRESSO} ab initio package
(MLWFs have been generated as a post-processing calculation using
the WanT package\cite{WanT}): we modeled the substrate 
using a periodically-repeated hexagonal supercell,
with a $(\sqrt{3}\times \sqrt{3})R30^{\circ}$ structure and a surface slab
made of 15 Cu atoms distributed over 5 layers;
the Brillouin Zone has been sampled using a 
$6\times6\times1$ $k$-point mesh.

By fitting the adatom binding energy as a function of its distance
from the substrate, $z$,
(as it is usually done\cite{DaSilva05} the fit has been
performed by optimizing the
parameters of the function: $A\,e^{-Bz}-C_3/(z-z_0)^3$ ), one can easily
estimate the $C_3$ coefficients that characterize the adatom-surface
vdW interactions.
As can be seen in Table II, although the agreement with reference 
$C_3$ data is not perfect yet (obtaining accurate $C_3$ coefficients 
represents a notorious difficult problem, see, for instance,
ref. \onlinecite{Bruch}), nonetheless the DFT/vdW-WF2 method 
gives a dramatic improvement with respect to the previous DFT/vdW-WF scheme.

In conclusion, we have described and applied a new implementation
of our vdW-correction method based on the Maximally-Localized Wannier
functions: the DFT/vdW-WF2 approach is based on the London expression 
and takes into account the MLWF intrafragment overlap.
The application to selected dimers and also to
Ar on graphite and Ne on the Cu(111) metal surface
show a substantial improvement in the long-range
vdW-coefficient ($C_6$ and $C_3$) estimates.
Work is in progress to achieve a similar level of
improvement in equilibrium distances
and binding energies: this would probably require the introduction of 
more sophisticated, DFT-functional dependent, damping functions.

\begin{table}[ht]
\begin{tabular}{lrrrrrr}
\multicolumn{7}{c}{} \\
\hline
\hline
		  	&   &DFT/vdW-WF  &  &DFT/vdW-WF2 &   & Ref.     \\
\hline
H-H	 		&   & 7.50(8.0)	 &  & 7.17(7.48) &   & 6.38 	\\
He-He	 		&   & 0.57(0.62) &  & 1.48(1.47) &   & 1.45 	\\
Ne-Ne	 		&   & 4.35(4.73) &  & 10.4(8.9)  &   & 6.35 	\\
Ne-Ar	 		&   & 24.9(16.9) &  & 26.4(22.9) &   & 19.5     \\
Ar-Ar	 		&   & 92.5(93.2) &  & 65.8(66.1) &   & 64.3     \\
Kr-Kr	 		&   &214.0(227.0)&  &124.0(124.0)&   & 131.0	\\
Xe-Xe	 		&   &618.0(621.0)&  &261.0(262.0)&   & 285.9	\\
N2-N2	 		&   & 87.4(89.3) &  & 81.2(80.5) &   & 73.3 	\\
CO-CO	 		&   & 85.6(86.7) &  & 84.8(85.1) &   & 81.5  	\\
NH$_3$-NH$_3$	 	&   & 67.1(88.4) &  & 63.5(77.6) &   & 89.03 	\\
H$_2$O-H$_2$O	 	&   & 35.2(35.6) &  & 38.9(37.3) &   & 45.29 	\\
C$_2$H$_6$-C$_2$H$_6$ 	&   &308.0(315.0)&  &298.0(300.0)&   & 381.9 	\\
CH$_4$-CH$_4$ 		&   &103.0(119.0)&  & 98.2(111.0)&   & 129.7 	\\
C$_6$H$_6$-C$_6$H$_6$ 	&   &2930.0(2900)&  &1710.0(1710)&   & 1722.7  	\\
C$_6$H$_6$-Ar 		&   &490.0(495.0)&  &333.0(334.0)&   & 330.1  	\\
C$_6$H$_6$-H$_2$O 	&   &323.0(325.0)&  &252.0(256.0)&   & 277.4 	\\
CO$_2$-CO$_2$	 	&   &187.0(191.0)&  &162.0(158.0)&   & 158.5   	\\
NH$_3$-CO	 	&   & 78.5(88.6) &  & 75.5(80.4) &   & 90.2   \\
\hline
MRE			&   &14.1(16.6)\%&  &0.3(-0.3)\% &   &       \\	
\hline
MARE			&   &35.4(32.8)\%&  &14.6(10.8)\%&   &       \\			
\hline
\hline
\end{tabular}
\caption{$C_6$ coefficients(in meV \AA$^6$), 
using the reference DFT revPBE functional
(PBE in parenthesis) computed with the DFT/vdW-WF2 method,
compared with those obtained by the previous 
DFT/vdW-WF scheme,\cite{silvprl} and with reference values.\cite{52,54,55,18,molphy,t1,t4,t5,t13}}
\label{tabellac6}
\end{table}

\begin{table}[ht]
\begin{tabular}{lrrrrrr}
\multicolumn{7}{c}{} \\
\hline
\hline
		  	&   &DFT/vdW-WF  &  &DFT/vdW-WF2 &   & Ref.     \\
\hline
Ar-graphite             &   & 18318      &   & 2057      &   & 1210 \\
Ne-Cu(111)              &   &  1226      &   &  589      &   &  488 \\    
\hline
\hline
\end{tabular}
\caption{$C_3$ coefficients (in meV \AA$^3$), 
using the reference DFT revPBE functional computed with the DFT/vdW-WF2 method,
compared with those obtained by the previous 
DFT/vdW-WF scheme,\cite{silvprl} and with reference values.\cite{Vidali}}
\label{tabellac3}
\end{table}

\end{document}